\documentclass[twocolumn,prd,amsmath,amssymb,floatfix]{revtex4}

\usepackage{bbold}
\usepackage{bm}
\usepackage{dashbox}
\usepackage{color}
\usepackage{dcolumn}
\usepackage{feynmf} 
\usepackage{graphics}
\usepackage{graphicx}
\usepackage{subfigure}
\usepackage{slashed}
\usepackage{placeins}

\allowdisplaybreaks
\usepackage{pdfpages}
\usepackage{eso-pic}
\usepackage{everyshi}
\usepackage{pdflscape}

\newcommand{\ep}{\varepsilon}

\def\be{\begin{equation}}
\def\ee{\end{equation}}
\def\bea{\begin{eqnarray}}
\def\eea{\end{eqnarray}}
\def\bse{\begin{subequations}}
\def\ese{\end{subequations}}
\def\bc{\begin{center}}
\def\ec{\end{center}}

\def\nonum{\nonumber}

\def\I{{\rm i}}

\def\D{{\rm d}}
\def\Ord{{\rm O}}

\newcommand{\ie}{{\it i.e.}}
\newcommand{\eg}{{\it e.g.}}

\begin{document}

\title{Landau-Khalatnikov-Fradkin transformation and \\ the mystery of even $\zeta$-values in Euclidean massless correlators}
       \author{
       A.\ V.~Kotikov$^{1}$ and S.~Teber$^{2}$}
\affiliation{
$^1$Bogoliubov Laboratory of Theoretical Physics, Joint Institute for Nuclear Research, 141980 Dubna, Russia.\\
$^2$Sorbonne Universit\'e, CNRS, Laboratoire de Physique Th\'eorique et Hautes Energies, LPTHE, F-75005 Paris, France.
 }

\date{\today}

\begin{abstract}
	The Landau-Khalatnikov-Fradkin (LKF) transformation is a powerful and elegant transformation allowing to study the
	gauge dependence of the propagator of charged particles interacting with gauge fields. With the help of this transformation, we
	derive a non-perturbative identity between massless propagators in two different gauges. From this identity, we 
	find that the corresponding perturbative series can be exactly expressed in terms of a hatted transcendental basis
	that eliminates all even $\zeta$-values. This explains the mystery of even $\zeta$-values observed 
	in multi-loop calculations of Euclidean massless correlators for almost three decades now. 
	Our construction further allows us to derive an exact formula relating hatted 
	and standard $\zeta$-values to all orders of perturbation theory. 
  
\end{abstract}

\maketitle

\section{Introduction }

Gauge invariance governs the dynamics of systems of charged particles with deep consequences in elementary particle physics and beyond. 
Through the gauge principle, it gives rise to gauge field theories the prototype of which is quantum electrodynamics (QED). 
In the latter, gauge freedom can be expressed via a covariant gauge fixing procedure that introduces an explicit 
dependence of the Lagrangian on a gauge fixing parameter $\xi$.  While physical quantities should not depend on this parameter, 
precious information can be obtained by studying the $\xi$-dependence of various correlation functions. 

Such a task can be carried out with the help of the Landau-Khalatnikov-Fradkin (LKF) 
transformation~\cite{Landau:1955zz} (see also~[\onlinecite{Johnson:1959zz,Sonoda:2000kn}]) 
that elegantly relates the QED fermion propagator in two different $\xi$-gauges (and similarly for the fermion-photon vertex). 
Its most important applications~\cite{Curtis:1990zs,Burden:1998gr,Jia:2016udu} are related to the study of the gauge covariance of QED Schwinger-Dyson 
equations and their solutions. This allows, \eg, to construct a charged-particle-photon vertex ansatz both in 
scalar \cite{Fernandez-Rangel:2016zac} and spinor QED \cite{Kizilersu:2009kg}. 
Other applications~\cite{Bashir:2002sp,Jia:2016wyu} are focused on estimating large orders of perturbation theory. 
Indeed, and this will play a crucial role in what follows, the non-perturbative nature of the LKF transformation allows to fix some of the
coefficients of the all-order expansion of the fermion propagator. Starting with a perturbative propagator in some fixed gauge, say $\eta$, all
the coefficients depending on the difference between the gauge fixing parameters of the two propagators, $\xi - \eta$, get fixed by a weak coupling expansion of 
the LKF-transformed initial one. Such estimations have been carried out for QED in various dimensions (see [\onlinecite{Bashir:2002sp,Jia:2016wyu}]), for
generalizations to brane worlds~\cite{Ahmad:2016dsb} and for more general SU(N) gauge theories~\cite{DeMeerleer:2018txc}.


A seemingly unrelated topic is focused on the multi-loop structure of propagator-type functions 
(p-functions~\footnote{Following~\cite{Baikov:2018gap}, by p-functions we understand ($\overline{\text{MS}}$-renormalized) 
Euclidean 2-point functions (that can also be obtained from 3-point functions by setting one external momentum to 
zero with the help of infra-red rearrangement) expressible in terms of massless propagator-type Feynman integrals also known as p-integrals.}). 
About three decades ago, it was noticed that all contributions proportional to $\zeta_4 = \pi^4/90$ mysteriously cancel out in the Adler 
function at three-loops~\cite{Gorishnii:1990vf}. Two decades later, it was shown that the four-loop contribution is also $\pi$-free and 
that a similar fact holds for the coefficient function of the Bjorken sum rule~\cite{Baikov:2010je}. There is by now mounting 
evidence, see, \eg, \cite{Baikov:2016tgj,Herzog:2017ohr,Davies:2017hyl,Moch:2018wjh}, that various massless Euclidean physical quantities demonstrate striking regularities in terms proportional 
to even $\zeta$-function values, $\zeta_{2n}$, {\it e.g.}, to $\pi^{2n}$ with $n$ being a positive integer 
\footnote{Notice also that, within a Schwinger-Dyson equation approach in fixed dimension, renormalized Euclidean massless correlators were shown to be expressed only 
in terms of odd zeta-values \cite{Kreimer:2006ua}.}. 
Additional cancellations of $\pi^{2n}$ terms have been observed in the so-called C-scheme \cite{Jamin:2017mul}.
Such puzzling facts have recently given rise to the ``no-$\pi$ theorem''. The latter is based on the 
 observation~\cite{Broadhurst:1999xk,Baikov:2010hf} that the $\ep$-dependent transformation of the $\zeta$-values:
\be
\hat{\zeta}_3 \equiv \zeta_3 + \frac{3\ep}{2} \zeta_4 - \frac{5\ep^3}{2} \zeta_6,~~ \hat{\zeta}_5 \equiv \zeta_5 + \frac{5\ep}{2} \zeta_6,~~
\hat{\zeta}_7 \equiv \zeta_7\, ,
\label{hatZe}
\ee
eliminates even zetas from the expansion of four-loop p-integrals. A generalization of (\ref{hatZe}) to 5- and 6-loops is available
in Refs.~[\onlinecite{Baikov:2018wgs,Georgoudis:2018olj}] and [\onlinecite{Baikov:2018gap}], respectively (the latter used the results obtained in \cite{Lee:2011jt})~\footnote{After 
the submission of our paper, a new paper \cite{Baikov:2019zmy} appeared on the ArXiv, where the results (\ref{hatZe}) have been extended to 7 loop order.}.
The results (\ref{hatZe}) and their extensions in \cite{Georgoudis:2018olj,Baikov:2018gap,Baikov:2019zmy} give a possibility to predict
  the terms $\sim \pi^{2n}$ in higher orders of perturbation theory (see their evaluation in \cite{Baikov:2018wgs,Baikov:2018gap,Baikov:2019zmy}).
Note that the results \cite{Baikov:2018wgs,Georgoudis:2018olj,Baikov:2018gap,Baikov:2019zmy}
  also contain multi-zeta values the consideration of which
is beyond the scope of the present study.

In the present paper, we shall use the LKF transformation in order to study general properties of the coefficients of the propagator. 
We will show how the transformation naturally reveals the existence of the hatted transcendental basis. Moreover, it will allow
us to extend the results of Eq.~(\ref{hatZe}) to any order in $\ep$.

 The appearance of the hatted transcendental basis from the LKF transformation can be naturally understood in the following way.
  The LKF transformation produces all-loop results for very restricted objects: the difference of fermion propagators in two gauges.
  So, at every order of the  $\ep$-expansion
  these all-loop results should contain 
  (at least, a part of) the basic properties of
  the corresponding master integrals, \ie, the all-loop results should be expressed in the form of (at least, a part of)
  the  corresponding hatted $\zeta$-values. In a sense, it is not the full set
  of the hatted $\zeta$-values but only the one-fold ones. This comes from the fact that the results produced by the LKF transformation
  contain only products of $\Gamma$-functions and, thus, their expansions contain only the simple one-fold $\zeta$-values.

The paper is organized as follows.  Sec.~\ref{sec:LKF:x-space} introduces the LKF transformation in coordinate space.
The corresponding results in momentum space are presented in Secs.~\ref{sec:LKF:p-space:F} and \ref{sec:LKF:p-space:S} for spinor QED and its scalar counter-part, respectively.
Sec.~\ref{sec:LKF:proof} then provides the evaluation of the basic elements of the LKF transformation in the form of hatted $\zeta$-values.
The results are summarized in Sec.~\ref{sec:summary} and some of the first coefficients of the expansion of the hatted $\zeta$-values in terms of the usual ones are
displayed in App.~\ref{app:sec:express}.

\section{LKF transformation}
\label{sec:LKF:x-space}

In the following, we shall consider QED in an Euclidean space of dimension $d$ ($d=4-2\ep$).
%
%
The general form of the fermion propagator $S_F(p,\xi)$ in some gauge $\xi$ reads:
\be
S_F(p,\xi) = \frac{1}{i\hat{p}} \, P(p,\xi) \, ,
\label{SFp}
\ee
where the tensorial structure, \eg, the factor $\hat{p}$ containing Dirac $\gamma$-matrices, has been extracted.
It is also convenient to introduce the $x$-space representation $S_F(x,\xi)$
of the fermion propagator as:
\be
S_F(x,\xi) =  \hat{x} \, X(x,\xi) \, .
\label{SFx}
\ee
The two representations, $S_F(x,\xi)$ and $S_F(p,\xi)$, are related by the Fourier transform which is defined as:
\begin{subequations}
\label{Fourier:def}
\begin{flalign}
&S_F(p,\xi) = \int \frac{\D^dx}{(2\pi)^{d/2} } \, e^{\I px} \, S_F(x,\xi) \, , 
\label{SFx2p} \\
&S_F(x,\xi) = \int \frac{\D^dp}{(2\pi)^{d/2} } \, e^{-\I px} \, S_F(p,\xi) \, .
\label{SFp2x}
\end{flalign}
\end{subequations}

The famous LKF transformation connects in a very simple way the fermion propagator in two different gauges, \eg, $\xi$ and $\eta$. In dimensional regularization, it reads:
%
\be
S_F(x,\xi) = S_F(x,\eta) \, e^{\I\,(D(x)-D(0))} \, ,
\label{LKF} 
\ee
where
\be
D(x)= -\I\,\Delta\, e^2\, \mu^{4-d} \int \frac{\D^dp}{(2\pi)^{d} } \, \frac{e^{-\I p x}}{p^4},~~ \Delta=\xi-\eta \, .
\label{Dx}
\ee
%
Note that, in dimensional regularization, the term $D(0)$ is proportional to the massless tadpole 
 $T_2$, the massive counterpart of which is defined as:
\be
T_{\alpha}(m^2) =  \int \frac{\D^dp}{(2\pi)^{d} } \, \frac{e^{-\I px}}{(p^{2}+m^2)^{\alpha}} \, .
\label{Talpha}
\ee
The tadpole $T_{\alpha}(m^2) \sim \delta(\alpha-d/2)$ in the massless limit and, thus, $D(0)=0$ in the framework of dimensional regularization. 
So, Eq.~(\ref{LKF}) can be simplified as follows:
\be
S_F(x,\xi) = S_F(x,\eta)\, e^{\I D(x)} \, .
\label{LKFN}
\ee

We may now proceed in calculating $D(x)$. In order to do so, it is possible to use the following simple formulas
for the Fourier transform of massless propagators (see, for example, \cite{Kotikov:2018wxe}):
\begin{subequations}
\label{Fourier:formulas}
\begin{flalign}
&\int \D^dx \, \frac{e^{\I px}}{x^{2\alpha}} \, = \frac{2^{2\tilde{\alpha}} \pi^{d/2} a(\alpha)}{p^{2\tilde{\alpha}}},~~
  a(\alpha)=\frac{\Gamma(\tilde{\alpha})}{\Gamma(\alpha)},~~\tilde{\alpha}=\frac{d}{2}-\alpha \, ,
\label{SFx2pN}\\
&\int \D^dp \, \frac{e^{-\I px}}{p^{2\alpha}} \, = \frac{2^{2\tilde{\alpha}} \pi^{d/2} a(\alpha)}{x^{2\tilde{\alpha}}}\, .
\label{SFp2xN}
\end{flalign}
\end{subequations}
This yields:
\be
D(x)=-\I \, \Delta \,e^2\, (\mu^2 x^2)^{2-d/2} \, \frac{\Gamma(d/2-2)}{2^4(\pi)^{d/2}}\, ,
\label{DxN0}
\ee
or, equivalently, with the parameter $\ep$ made explicit:
\begin{flalign}
D(x)=\frac{\I\,\Delta\,A}{\ep}\,\Gamma(1-\ep)\, (\pi\mu^2 x^2)^{\ep} , ~~ A=\frac{\alpha_{\rm em}}{4\pi}=\frac{e^2}{(4\pi)^2}\, .
\label{DxN}
\end{flalign}
From Eq.~(\ref{DxN}), we see that $D(x)$ contributes with a common factor $\Delta A$ accompanied by the singularity $\ep^{-1}$.

Note that, in the case of the scalar QED, the corresponding scalar propagator $S_C(x,\xi)$  has a similar transformation
between $\xi$ and $\eta$ gauges, \ie,
\be
S_C(x,\xi) = S_C(x,\eta)\, e^{\I D(x)} \, .
\label{LKFN.c}
\ee

\section{LKF transformation in momentum space}
\label{sec:LKF:p-space:F}

Let's assume that, for some gauge fixing parameter $\eta$,
the fermion propagator $S_F(p,\eta)$ with external momentum $p$ has the form (\ref{SFp}) with $P(p,\eta)$ reading:
\be
P(p,\eta) = \sum_{m=0}^{\infty} a_m(\eta)\, A^m \,{\left(\frac{\tilde{\mu}^2}{p^2}\right)}^{m\ep} \, .
\label{Peta}
\ee
In Eq.~(\ref{Peta}), the $a_m(\eta)$ are coefficients of the loop expansion of the propagator and
$\tilde{\mu}$ is the renormalization scale:
\be
\tilde{\mu}^2= 4\pi \mu^2 \, ,
\label{Aem}
\ee
which lies somehow between the MS-scale $\mu$ and the $\overline{\text{MS}}$-scale $\overline{\mu}$.
Then, the LKF transformation shows that, for another gauge parameter $\xi$,
the fermion propagator can be expressed as:
%
\be
\hspace{-5pt}P(p,\xi) = \sum_{m=0}^{\infty} a_m(\xi)\, A^m\, {\left(\frac{\tilde{\mu}^2}{p^2}\right)}^{m\ep} \, ,
\label{Pxi} 
\ee
where
\begin{flalign}
&\hspace{-5pt}a_m(\xi) = a_m(\eta) \, \frac{\Gamma(2-(m+1)\ep)}{\Gamma(1+m\ep)}\, \times 
	\nonum \\
&\hspace{-5pt}\times\,\sum_{l=0}^{\infty} 
\frac{\Gamma(1+(m+l)\ep)\,\Gamma^l(1-\ep)}{l!\,\Gamma(2-(m+l+1)\ep)} \, \frac{(\Delta \, A)^l}{(-\ep)^l} \, {\left(\frac{\tilde{\mu}^2}{p^2}\right)}^{l\ep} \, .
\label{axi}
\end{flalign}
In order to derive (\ref{axi}), we used the fermion propagator $S_F(p,\eta)$ with  $P(p,\eta)$ given by (\ref{Peta}), did the Fourier transform to
$S_F(x,\eta)$ and applied the LKF transformation (\ref{LKFN}). As a final step, we took the inverse Fourier transform and obtained 
$S_F(p,\xi)$ with  $P(p,\xi)$ given by (\ref{Pxi}).

\subsection{Scale fixing}

In our present study, we consider only the case of the so-called MS-like schemes (see, for example, $\tilde{\mu}$ above in (\ref{Aem})).
In such schemes, we need to fix specific terms coming from the application of dimensional regularization. Such a procedure will be called
{\it scale fixing} 
and will play a crucial role in our analysis.
	
Let's first recall that the $\overline{\text{MS}}$-scale $\overline{\mu}$ is related to the previously defined scale $\tilde{\mu}$ with the help of:
%
%
\be
\overline{\mu}^2=\tilde{\mu}^2\, e^{-\gamma},
\label{Delta}
\ee
where $\gamma$ is the Euler constant. An advantage of the $\overline{\text{MS}}$-scale is that it subtracts the Euler constant $\gamma$ from the $\ep$-expansion.
Moreover, it is well known that, in calculations of two-point massless diagrams, the final results do not display any $\zeta_2$
~\footnote{Strictly speaking, $\zeta_2$ can appear in some formulas such as sum rules in deep-inelastic scattering. They originate from 
an analytic continuation \cite{Kotikov:2005gr} of certain special forms of p-integrals. We will not consider this case in the present study.}.
 So it is convenient to choose some scale which also subtracts $\zeta_2$ in intermediate steps of the calculation.
For this purpose, we shall consider two different scales. 
	
The first one is the popular $G$-scale \cite{Chetyrkin:1980pr}, which subtracts the coefficient in factor of
the singularity $1/\ep$ in the one-loop scalar p-type integral, \ie,
\be
\mu_G^{2\ep}=\tilde{\mu}^{2\ep} \, \frac{\Gamma^2(1-\ep)\Gamma(1+\ep)}{\Gamma(2-2\ep)}\, .
\label{G-scale}
\ee
Following \cite{Broadhurst:1999xk}, we shall use a slight modification of this scale that we will refer to as the $g$-scale
and in which an additional factor $1/(1-2\ep)$ is subtracted from the one-loop result, \ie,
\be
\mu_g^{2\ep}=\tilde{\mu}^{2\ep} \, \frac{\Gamma^2(1-\ep)\Gamma(1+\ep)}{\Gamma(1-2\ep)}\, .
\label{g-scale}
\ee
The advantage of the $g$-scale (over the $G$-scale) will reveal itself in discussions below related to the so-called transcendental weight of various contributions. 

We shall also introduce a new scale which is based on old calculations of massless diagrams performed by Vladimirov who added \cite{Vladimirov:1979zm}
an additional factor $\Gamma(1-\ep)$ to each loop contribution. The latter corresponds to adding the factor $\Gamma^{-1}(1-\ep)$ to the corresponding scale. 
We shall refer to this scale as the minimal Vladimirov-scale, or MV-scale, and define~\footnote{Notice that the form (\ref{Vla}) has been used once to define the
$\overline{\text{MS}}$ scheme (see Errata to Ref.~\cite{Kataev:1988sq}).}:
%
\be
\mu_{\text{MV}}^{2\ep}=\frac{\tilde{\mu}^{2\ep}}{\Gamma(1-\ep)} \, .
\label{Vla}
\ee
As we will show below, the use of the MV-scale leads to simpler results in comparison with the $g$ one. Hence, the MV-scale is more appropriate to our analysis and 
all our basic results will be given in the MV-scale. After that we will discuss the differences coming from the use of the $g$-scale. 

%
In both the MV-scale and $g$-scale,
we can rewrite the result (\ref{axi}) in the following general form:
\begin{flalign}
a_m(\xi) &= a_m(\eta)
\sum_{l=0}^{\infty} \, \frac{1-(m+1)\ep}{1-(m+l+1)\ep} \, \times
\nonum \\
&\times 
\, \Phi_p(m,l,\ep) 
\, \frac{(\Delta \, A)^l}{(-\ep)^l l!} \, {\left(\frac{\mu_p^2}{p^2}\right)}^{l\ep} \, ,
\label{axi.1}
\end{flalign}
where $p=$MV$,g$ see also Sec.~\ref{sec:LKF:p-space:S} for very similar results in the case of scalar QED. 
In Eq.~(\ref{axi.1}), the factor $(1-(m+1)\ep)/(1-(m+l+1)\ep)$ has been specially extracted from $\Phi_p(m,l,\ep)$ in order to insure 
equal transcendental level, \ie, the same value of $s$ for $\zeta_s$ at every order of the $\ep$-expansion of $\Phi_p(m,l,\ep)$ (see below). 
Central to the present work, the factors $\Phi_{\text{MV}}(m,l,\ep)$ and  $\Phi_g(m,l,\ep)$ read:
\begin{subequations}
\label{Phi:V+g}
\begin{flalign}
&\Phi_{\text{MV}}(m,l,\ep)=\frac{\Gamma(1-(m+1)\ep)\Gamma(1+(m+l)\ep)\Gamma^{2l}(1-\ep)}{
 \Gamma(1+m\ep)\Gamma(1-(m+l+1)\ep)} \, ,
 \label{Phi:V:def} \\
&\Phi_g(m,l,\ep)=\Phi_{\text{MV}}(m,l,\ep) \, \frac{\Gamma^l(1-2\ep)}{\Gamma^{3l}(1-\ep)\Gamma^l(1+\ep)}\, ,
\label{Phi:g:def}
\end{flalign}
\end{subequations}
and may be expressed as expansions in $\zeta_i$ $(i\geq 3)$ that we now proceed on studying.
%

\subsection{MV-scale}

The $\Gamma$-function $\Gamma(1+\beta\ep)$ has the following expansion:
\begin{flalign}
\Gamma(1+\beta\ep) = \exp \Big[ -\gamma \beta \ep + \sum_{s=2}^{\infty}\, (-1)^s \, \eta_s \beta^s \ep^s \Bigr],~~ 
\eta_s = \frac{\zeta_s}{s} \, .
\label{Gamma:exp}
\end{flalign}
Substituting Eq.~(\ref{Gamma:exp}) in Eq.~(\ref{Phi:V:def}), yields for the factor $\Phi_{\text{MV}}(m,l,\ep)$:
\be
\Phi_{\text{MV}}(m,l,\ep)= \exp \Big[ \sum_{s=2}^{\infty}\,\eta_s \, p_s(m,l) \, \ep^s \Bigr]\, ,
\label{Phi:V}
\ee
where
\begin{flalign}
p_s(m,l)= &~ (m+1)^s-(m+l+1)^s + 2l + 
\nonum \\
& + (-1)^s \Bigl\{(m+l)^s-m^s\Bigr\}\, ,
\label{ps:V}
\end{flalign}
and, as expected from the MV-scale, we do have:
\be
p_1(m,l)=0,~~ p_2(m,l)=0 \, .
\label{ps:V:1-2}
\ee



As can be see from Eq.~(\ref{Phi:V}), $\Phi_{\text{MV}}(m,l,\ep)$ contains $\zeta_s$-function values of a given weight (or transcendental level) $s$ in factor of $\ep^s$. 
Such a property strongly constrains the coefficients of the $\ep$-series thereby simplifying our analysis. It is reminiscent of the one earlier found in Ref.~\cite{Kotikov:2000pm}.
When judiciously used, it sometimes allows to derive results without any calculations (as in Ref.~\cite{Kotikov:2002ab}).
In other cases, it simplifies the structure of the results which can then be predicted as an ansatz in a very simple way
(see Refs.~\cite{Fleischer:1998nb,Kotikov:2007cy}). For a recent application of such property, see the recent papers \cite{Dixon:2019uzg} and references and discussions therein.

\section{Scalar QED}
\label{sec:LKF:p-space:S}


In the case of scalar (spin-$0$) QED, the LKF transformation leads to expressions which are very similar to spinor QED that we have considered so far.

Indeed, let $S_C(p,\eta)$ be a scalar propagator with external momentum $p$ and gauge fixing parameter $\eta$; its general expression reads:
%
\be
S_C(p,\eta) = \frac{1}{p^2} \, \sum_{m=0}^{\infty} a^c_m(\eta)\, A^m \,{\left(\frac{\tilde{\mu}^2}{p^2}\right)}^{m\ep} \, ,
\label{Sc.eta}
\ee
where $a^c_m(\eta)$ are coefficients of the loop expansion of the propagator and
$\tilde{\mu}$ the renormalization scale (\ref{Aem}).
Then, the LKF transformation shows that, for another gauge parameter $\xi$,
the scalar propagator can be expressed as:
\be
S_C(p,\xi) = \frac{1}{p^2} \, \sum_{m=0}^{\infty} a^c_m(\xi)\, A^m \,{\left(\frac{\tilde{\mu}^2}{p^2}\right)}^{m\ep} \, ,
\label{Sc.xi}
\ee
where
\begin{flalign}
a^c_m(\xi) &= a^c_m(\eta)\, \frac{\Gamma(1-(m+1)\ep)}{\Gamma(1+m\ep)} \, \times
\nonum \\
&\times 
\, \sum_{l=0}^{\infty}
\frac{\Gamma(1+(m+l)\ep)\,\Gamma^l(1-\ep)}{l!\,\Gamma(1-(m+l+1)\ep)} \, \frac{(\Delta \, A)^l}{(-\ep)^l} \, {\left(\frac{\tilde{\mu}^2}{p^2}\right)}^{l\ep}\, .
\label{axi.c}
\end{flalign}
In order to derive (\ref{axi.c}), we used the scalar propagator $S_C(p,\eta)$ of Eq.~(\ref{Sc.eta}), did the Fourier
transform to $S_C(x,\eta)$ and applied the LKF transformation (\ref{LKFN.c})
%
%
which is identical to (\ref{LKFN}). As a final step, we took the inverse Fourier transform and obtained
$S_C(p,\xi)$ in (\ref{Sc.xi}).

For the MV- and $g$-scales,
we can rewrite Eq.~(\ref{axi.c}) in the following form:
\be
a^c_m(\xi) = a^c_m(\eta) \,
\sum_{l=0}^{\infty} \,
\Phi_p(m,l,\ep)
\, \frac{(\Delta \, A)^l}{(-\ep)^l l!} \, {\left(\frac{\mu_p^2}{p^2}\right)}^{l\ep},
\label{axi.1c}
\ee
where $p=$MV$,g$. We therefore see that the difference between Eqs.~(\ref{axi.1c}) and (\ref{axi.1})
is only in the factor $(1-(m+1)\ep)/(1-(m+l+1)\ep)$ which did not play any role in the above analysis.
Hence, for scalar QED, we can repeat all evaluations done for spinor QED and we will end up with the same
set of hatted $\zeta$-values as before in Eqs.~(\ref{hZeta}) and (\ref{hC}).

\section{Solution of the recurrence relations}
\label{sec:LKF:proof}

We now focus on the polynomial $p_s(m,l)$ of Eq.~(\ref{ps:V}) that is conveniently separated in even and odd $s$ values. Then, we see that the following recursion relations hold:
\begin{subequations}
\label{recursions:V}
\begin{flalign}
	p_{2k} &= p_{2k-1} + L p_{2k-2} + p_{3}, \quad L=l(l+1) \, , 
\label{p2k:V} \\
	p_{2k-1} &= p_{2k-2} + L p_{2k-3} + p_{3} \, .
\label{p2k-1:V}
\end{flalign}
\end{subequations}
Specific to the MV-scheme, these relations only depend on $L$ which leads to strong simplifications. Nevertheless, they are difficult to solve for arbitrary $k$. It is simpler 
to proceed by explicitly considering the first values of $k$:
%
\begin{subequations}
\label{ps:V:values}
\begin{flalign}
&p_4= 2p_3\, , 
\label{p4:V}\\
&p_5= p_4 + Lp_3 + p_3 = (3+L)p_3\, , 
\label{p5:V}\\
&p_6= p_5 + Lp_4 + p_3 = (4+3L)p_3\, ,
\label{p6:V}
\end{flalign}
\end{subequations}
showing that $p_s$ takes the form of a polynomial in $L$ in factor of $p_3$.
Then, taking $Lp_3$ from (\ref{p5:V}) and substituting it in (\ref{p6:V}) yields:
\be
Lp_3=p_5-3p_3, \quad p_6=3p_5-5p_3 \, ,
\label{p6:V:express}
\ee
which reveals that the even polynomial $p_6$ can be entirely expressed in terms of the lower order odd ones, $p_3$ and $p_5$.

We may automate this procedure for higher values of $k$. The general expression of $p_s$ is given by: 
\be
p_s=\sum_{m=0}^{[\frac{s+1}{2}-2]} \, A_{s,m} \, L^m \, p_3 \, ,
\label{ps:V:general}
\ee
see Eqs.~(\ref{appA:Akm:V:values}) of App.~\ref{app:sec:express} for values of the first 20 coefficients and Eqs.~(\ref{appA:ps:V:values})
for values of $p_s$ with $7 \leq s \leq 20$. Taking $L^k p_3$ from the equations for $p_{2k-1}$ and substituting them in the equations for $p_{2k}$ yields:
%
\begin{flalign}
p_{2k}=\sum_{s=2}^{k} p_{2s-1} \, C_{2k,2s-1} \, =  \sum_{m=1}^{k-1} p_{2k-2m+1} \, C_{2k,2k-2m+1} \, ,
\label{p2k:V:general}
\end{flalign}
where the 20 first coefficients, $C_{2k,2s-1}$, are presented in App.~\ref{app:sec:express}, Eq.~(\ref{appA:C:V:values}).
From these results, it is possible to determine the exact $k$-dependence
of $C_{2k,2s-1}$, which has the following structure:
\be
C_{2k,2k-2m+1} = b_{2m-1}
\, \frac{(2k)!}{(2m-1)! \, (2k-2m+1)!} \, ,
\label{C:V:structure}
\ee
with the first coefficients $b_{2m-1}$ taking the values:
\begin{flalign}
&b_{1}= \frac{1}{2},~~~~ b_{3}= -\frac{1}{4},~~~~ b_{5}= \frac{1}{2},~~~~ b_{7}= -\frac{17}{2},~~~~ b_{9}= \frac{31}{2}\, ,
\nonumber \\
&b_{11}= -\frac{691}{4},~~~~~ b_{13}= \frac{5461}{2},~~~~~ b_{15}= -\frac{929569}{16}\, ,
\nonumber \\	
&b_{17}= \frac{3202291}{2},~~~~~~ b_{19}= -\frac{221930581}{4} \, ,
\nonum \\ 
&b_{21}= \frac{4722116521}{2},~~~~~~ b_{23}= -\frac{968383680827}{8} \, .
\label{b:V:values}
\end{flalign}
Examining the numerators of $b_{2m-1}$, one can see that they are proportional to the numerators of Bernoulli
numbers. Indeed, a closer inspection reveals that, accurate to a sign, the coefficients $b_{2m-1}$
coincide with the zero values of Euler polynomials $E_n(x)$:
\be
b_{2m-1} \, = - E_{2m-1}(x=0)\, ,
\label{b:V:expressionE}
\ee
and therefore to Bernoulli and Genocchi numbers, $B_m$ and $G_m$, respectively, because
\begin{flalign}
E_{2m-1}(x=0) = \frac{G_{2m}}{2m}, \quad G_{2m} = - \frac{(2^{2m} - 1)}{m} \, B_{2m}\, .
\label{Ex=0}
\end{flalign}
Hence, the compact formula for the coefficients $b_{2m-1}$, expressed through the well known Bernoulli numbers $B_m$, reads:
\be
b_{2m-1} = \frac{(2^{2m} - 1)}{m} \, B_{2m} \, .
\label{b:V:expressionB}
\ee
Together with (\ref{C:V:structure}), Eq.~(\ref{b:V:expressionB}) provides an exact analytic expression for $p_{2k}$, Eq.~(\ref{p2k:V:general}), 
for arbitrary values of $k$.

\subsection{Hatted $\zeta$-values}


At this point, it is convenient to represent the argument of the exponential in the r.h.s.\ of (\ref{Phi:V}) as follows:
\begin{flalign}
\sum_{s=3}^{\infty}\,\eta_s \, p_s \, \ep^s = \sum_{k=2}^{\infty}\,\eta_{2k} \, p_{2k} \, \ep^{2k} +
\sum_{k=2}^{\infty}\,\eta_{2k-1} \, p_{2k-1} \, \ep^{2k-1} \, .
\label{Phi:V:exp}
\end{flalign}
%
%
With the help of Eq.~(\ref{p2k:V:general}), the first term in the r.h.s.\ of Eq.~(\ref{Phi:V:exp}) may be expressed as:
\begin{flalign}
\sum_{k=2}^{\infty}\,\eta_{2k} \, p_{2k} \, \ep^{2k} &= \sum_{k=2}^{\infty}\,\eta_{2k}  \, \ep^{2k} \, \sum_{s=2}^k p_{2s-1} \,
C_{2k,2s-1} 
\nonum \\
&= \sum_{s=2}^{\infty} p_{2s-1} \, \sum_{k=s}^{\infty}\,\eta_{2k} \, C_{2k,2s-1} \, \ep^{2k} \, .
\label{Phi:V:exp1}
\end{flalign}
Then, Eq.~(\ref{Phi:V:exp}) can be written as $\sum_{s=2}^{\infty}\,\hat{\eta}_{2s-1} \, p_{2s-1} \, \ep^{2s-1}$
%
%
where
\be
\hat{\eta}_{2s-1} = \eta_{2s-1} +  \sum_{k=s}^{\infty}\,\eta_{2k} \, C_{2k,2s-1} \, \ep^{2(k-s)+1} \, .
\label{hEta}
\ee
Thus, Eq.~(\ref{Phi:V}) can be represented as:
\begin{flalign}
\Phi_{\text{MV}}(m,l,\ep) &= \exp \Big[ \sum_{s=2}^{\infty}\,\hat{\eta}_{2s-1} \, p_{2s-1} \, \ep^{2s-1} \Bigr]
\nonum \\
&= \exp \Big[ \sum_{s=2}^{\infty}\,\frac{\hat{\zeta}_{2s-1}}{2s-1} \, p_{2s-1} \, \ep^{2s-1} \Bigr] \, ,
\label{Phi:V2}
\end{flalign}
where
\be
\hat{\zeta}_{2s-1} = \zeta_{2s-1} +  \sum_{k=s}^{\infty}\,\zeta_{2k} \, \hat{C}_{2k,2s-1} \, \ep^{2(k-s)+1} 
\label{hZeta}
\ee
with
\begin{flalign}
C_{2k,2s-1} &= b_{2k-2s+1} \, \frac{(2k)!}{(2s-1)! \, (2k-2s+1)!} \, , \nonumber \\ 
\hat{C}_{2k,2s-1} &= \frac{2s-1}{2k} \, C_{2k,2s-1}  
\nonum \\
&= b_{2k-2s+1} \, \frac{(2k-1)!}{(2s-2)! \, (2k-2s+1)!} 
\, .
\label{hC}
\end{flalign}
Together with (\ref{hC}) and (\ref{b:V:expressionB}), Eq.~(\ref{hZeta}) provides an exact expression for the hatted $\zeta$-values in terms of the standard
ones valid for all $\ep$, see Eqs.~(\ref{appA:hZeta:V:values}) for explicit expressions of $\hat{\zeta}_{2s-1}$ for $s \leq 10$.


\subsection{$g$-scale}

We may proceed in a similar way for the factor $\Phi_g(m,l,\ep)$. Substituting Eq.~(\ref{Gamma:exp}) in Eq.~(\ref{Phi:g:def}), yields:
\be
\Phi_g(m,l,\ep)= \exp \Big[ \sum_{s=2}^{\infty}\,\eta_s \, p^g_s(m,l) \, \ep^s \Bigr]\, ,
\label{Phi:g}
\ee
where the new polynomial $p^g_s(m,l)$ can be expressed in terms of $p_s(m,l)$, Eq.~(\ref{ps:V}), as:
\begin{flalign}
&p^g_s(m,l) = p_s(m,l) + \delta_s(m,l)\, , \nonumber \\
&\delta_s(m,l)= (2^s-3-(-1)^s)l \, .
\label{ps:g}
\end{flalign}
Eq.~(\ref{ps:g}) is such that $\delta_s(m,l)=0$ for $s=1$ and $s=2$ and, thus,
\be
p^g_1(m,l)=0, \quad p^g_2(m,l)=0\, ,
\label{ps:g:1-2}
\ee
similarly to the Vladimirov case, Eq.~(\ref{ps:V:1-2}).

We may then consider the even and odd values of $s$ separately as in (\ref{recursions:V}) leading to the following recursion relations:
\begin{subequations}
\label{recursions:g}
\begin{flalign}
	p^g_{2k} &= p_{2k} +  \delta_{2k}, \qquad ~~~ \delta_{2k}=4(2^{2k-2}-1)l \, ,
\label{p2k:g} \\
	p^g_{2k-1} &= p_{2k-1} + \delta_{2k-1}, \quad \delta_{2k-1}=\frac{1}{2} \delta_{2k}\, .
\label{p2k-1:g}
\end{flalign}
\end{subequations}
These recurrence relations depend on the variable $l$ but not on the product $L=l(l+1)$ as it was for the MV-scale. So, the $g$-scale
recursion relations (\ref{recursions:g}) are essentially more complicated than the MV-scale ones, (\ref{recursions:V}).
Fortunately, it is very simple to see that in the relations:
\be
p^g_{2k}=\sum_{s=2}^k p^g_{2s-1} \, C_{2k,2s-1} \, ,
\label{p2k:g:general}
\ee
the coefficients $C_{2k,2s-1}$ are exactly the same as in Eq.~(\ref{p2k:V:general}) because the corrections $\delta_{2k}$ and $\delta_{2k-1}$ 
exactly cancel each other. So, the hatted $\zeta$-values for the $g$-scale are identical to the ones of the MV-scale, (\ref{hZeta}) and (\ref{hC}).

\section{Summary}
\label{sec:summary}

From the result (\ref{axi.1}) corresponding to the LKF transformation of the fermion propagator (and similarly, from (\ref{axi.c}), 
corresponding to the LKF transformation of the scalar propagator) 
we have found peculiar recursion relations (\ref{p2k:V}) and (\ref{p2k-1:V}) between even and odd values of the polynomial associated to the 
uniformly transcendental factor $\Phi_{\text{MV}}(m,l,\ep)$ (\ref{Phi:V:def}). These relations are simple in the new MV-scheme
that we have introduced in Eq.~(\ref{Vla}). They relate the even and odd parts in a rather simple way (see Eq.~(\ref{p2k:V:general})) which reveals the possibility (\ref{Phi:V2})
to express all results for $\Phi_{\text{MV}}(m,l,\ep)$ in terms of hatted $\zeta$-values.
As can be seen from Eqs.~(\ref{p2k:g}) and (\ref{p2k-1:g}) in the more popular $g$-scheme, the corresponding recursion relations are
slightly more complicated but lead to the same relations (\ref{p2k:V:general}) between even and odd parts of the polynomial associated to 
$\Phi_g(m,l,\ep)$ (\ref{Phi:g:def}) and, correspondingly, to the same hatted $\zeta$-values. Our careful study of the recursion relations (\ref{p2k:V}) and (\ref{p2k-1:V})
allowed us to derive exact formulas, Eqs.~(\ref{hEta}) and (\ref{hZeta}), relating hatted and standard $\zeta$-values to all orders of perturbation theory. 
The coefficients of the relations are expressed trough the well-known Bernoulli numbers, $B_{2m}$, as can be see from Eqs.~(\ref{hC}) and (\ref{b:V:expressionB}). The numerical values of
some of these coefficients and some explicit relations between hatted and standard $\zeta$-values are presented in App.~\ref{app:sec:express}. Our results provide
stringent constraints on multi-loop calculations at any order in perturbation theory.

\acknowledgments

We are grateful to M.~Bellon, A.~Kataev, R.~Lee and S.~Mikhailov for valuable comments and discussions.

\begin{widetext}

\appendix

\section{Explicit expressions for $A_{k,m}$, $p_s$, $C_{2k,2s-1}$ and hatted $\zeta$-values}
\label{app:sec:express}

\subsection{Coefficients $A_{k,m}$ of Eq.~(\ref{ps:V:general}) for $k \leq 20$}

\begin{subequations}
\label{appA:Akm:V:values}
\begin{flalign}
&A_{k,0}=k-2,~~\,A_{2k-1,k-2}=1,~~\,A_{2k,k-2}=k \, ,
\\
&A_{7,1}=6, \qquad ~~ A_{8,1}=10, \quad~~~~~\,A_{9,1}=15, \quad~~~~~A_{9,2}=10, \quad~~~~\,A_{10,1}=21, \quad~~A_{10,2}=20 \, ,
\\
&A_{11,1}=28,~~~~~\,A_{11,2}=35, \quad~~~~A_{11,3}=15, \quad~~~~A_{12,1}=36, \quad~~~\,A_{12,2}=56, \quad~~A_{12,3}=35 \, ,
\\
&A_{13,1}=45,~~~~~\,A_{13,2}=84, \quad~~~~A_{13,3}=70, \quad~~~~A_{13,4}=21 \, ,
\\
&A_{14,1}=55,~~~~~\,A_{14,2}=120, \quad~~\,A_{14,3}=126, \quad~~\,A_{14,4}=56 \, ,
\\
&A_{15,1}=66,~~~~~\,A_{15,2}=165, \quad~~\,A_{15,3}=210, \quad~~\,A_{15,4}=126, \quad~~A_{15,5}=28 \, ,
\\
&A_{16,1}=78,~~~~~\,A_{16,2}=220, \quad~~\,A_{16,3}=330, \quad~~\,A_{16,4}=252, \quad~~A_{16,5}=84 \, ,
\\
&A_{17,1}=91,~~~~~\,A_{17,2}=286, \quad~~\,A_{17,3}=495, \quad~~\,A_{17,4}=462, \quad~~A_{17,5}=210, \quad~~A_{17,6}=36 \, ,
\\
&A_{18,1}=105,~~~~A_{18,2}=364, \quad~~\,A_{18,3}=715, \quad~~\,A_{18,4}=792, \quad ~~A_{18,5}=462, \quad~~A_{18,6}=120 \, ,
\\
&A_{19,1}=120,~~~~A_{19,2}=455, \quad~~\,A_{19,3}=1001, ~~~~A_{19,4}=1287, \quad \,A_{19,5}=924, \quad~~A_{19,6}=330,~~A_{19,7}=45 \, ,
\\
&A_{20,1}=136,~~~~A_{20,2}=560, \quad~~\,A_{20,3}=1365, ~~~~A_{20,4}=2002, \quad \,A_{20,5}=1716, \quad A_{20,6}=792,~~A_{20,7}=165 \, .
\end{flalign}
\end{subequations}

\subsection{Polynomials $p_s$ of Eq.~(\ref{ps:V:general}) for $7 \leq s \leq 20$}

%
\begin{subequations}
\label{appA:ps:V:values}
\begin{flalign}
&p_7= (5+6L+L^2)p_3 \, ,
\label{p7:V}\\
&p_8= (6+10L+4L^2)p_3 \, ,
\label{p8:V}\\
&p_9= (7+15L+10L^2+L^3)p_3 \, ,
\label{p9:V}\\
&p_{10}= (8+21L+20L^2+5L^3)p_3 \, ,
\label{p10:V}\\
&p_{11}= (9+28L+35L^2+15L^3+L^4)p_3 \, ,
\label{p11:V}\\
&p_{12}= (10+36L+56L^2+35L^3+6L^4)p_3 \, ,
\label{p12:V}\\
&p_{13}= (11+45L+84L^2+70L^3+21L^4+L^5)p_3 \, ,
\label{p13:V}\\
&p_{14}= (12+55L+120L^2+126L^3+56L^4+7L^5)p_3 \, ,
\label{p14:V}\\
&p_{15}= (13+66L+165L^2+210L^3+126L^4+28L^5+L^6)p_3 \, ,
\label{p15:V}\\
&p_{16}= (14+78L+220L^2+330L^3+252L^4+84L^5+8L^6)p_3 \, ,
\label{p16:V}\\
&p_{17}= (15+91L+286L^2+495L^3+462L^4+210L^5+36L^6+L^7)p_3 \, ,
\label{p17:V}\\
&p_{18}= (16+105L+364L^2+715L^3+792L^4+462L^5+120L^6+9L^7)p_3 \, ,
\label{p18:V}\\
&p_{19}= (17+120L+455L^2+1001L^3+1287L^4+924L^5+330L^6+45L^7+L^8)p_3 \, ,
\label{p19:V}\\
&p_{20}= (18+136L+560L^2+1365L^3+2002L^4+1716L^5+792L^6+165L^7+10L^8)p_3 \, .
\label{p20:V}
\end{flalign}
\end{subequations}

\subsection{Coefficients $C_{2k,2s-1}$ of Eq.~(\ref{p2k:V:general}) for $k \leq 10$}

\begin{subequations}
\label{appA:C:V:values}
\begin{flalign}
& C_{2k,2k-1}=k, \quad~~~~\,\, C_{6,3}=-5, \quad~~~~~~~ C_{8,5}=-14, \qquad~~~~\,\, C_{8,3}=28 \, ,
\\
& C_{10,7}=-30, \quad~~~~~\,\, C_{10,5}=126, \quad~~~~~ C_{10,3}=-255 \, ,
\\
& C_{12,9}=-55, \quad~~~~~\,\, C_{12,7}=396, \quad~~~~~ C_{12,5}=-1683, \quad~~~~ C_{12,3}=3410 \, ,
\\
& C_{14,11}=-91, \quad~~~~\,\, C_{14,9}=1001, \quad~~~\, C_{14,7}=-7293, \quad ~~~~ C_{14,5}=31031, \quad~~~\, C_{14,3}=-62881 \, ,
\\
& C_{16,13}=-140, \quad~~\,\,\, C_{16,11}=2184, ~~~~~\,C_{16,9}=-24310, ~~~~~\, C_{16,7}=177320, \quad~~ C_{16,5}=-754572 \, ,
\\
& C_{16,3}=1529080 \, ,
\\
& C_{18,15}=-204, \quad~~~\, C_{18,13}=4284, ~~~~~C_{18,11}=-67626, ~~~~\, C_{18,9}=753610, \quad ~~\, C_{18,7}=-5497596 \, ,
\\
& C_{18,5}=23394924,~~~C_{18,3}=-47408019 \, ,
\\
& C_{20,17}=-285, \quad~~~\,\, C_{20,15}=7752, ~~~~~C_{20,13}=-164730, ~~~~ C_{20,11}=2603380, ~~C_{20,9}=-29015090 \, ,
\\
& C_{20,7}=211668360,~~C_{20,5}=-900752361,~~C_{20,3}=1825305870 \, .
\end{flalign}
\end{subequations}

\subsection{Hatted $\zeta$-values, $\hat{\zeta}_{2s-1}$, of Eq.~(\ref{hZeta}) for $s \leq 10$}

We display below the first terms of the $\ep$-series generated by our exact result Eq.~(\ref{hZeta}). For the sake of clarity, we display
in a box 
the terms which were known up to the present work from the latest publication \cite{Baikov:2018gap}.  
Moreover, we also display in a dashed box 
 the terms calculated in the paper \cite{Baikov:2019zmy} which appeared in ArXiv two months after
the first version of our paper.

\begin{subequations}
\label{appA:hZeta:V:values}
\begin{flalign}
	\hat{\zeta}_{3} &= \boxed{\zeta_3 + \frac{3\,\ep}{2}\,\zeta_4 - \frac{5\,\ep^3}{2}\,\zeta_6 + \frac{21\,\ep^5}{2}\,\zeta_8 - \frac{153\,\ep^7}{2}\,\zeta_{10}}~ \scalebox{1}{\dbox{$+
	\frac{1705\,\ep^9}{2}\,\zeta_{12}$}} - \frac{26949\,\ep^{11}}{2}\,\zeta_{14} + \frac{573405\,\ep^{13}}{2}\,\zeta_{16} - \frac{15802673\,\ep^{15}}{2}\,\zeta_{18} +
\nonum \\
&+ \frac{547591761\,\ep^{17}}{2}\,\zeta_{20} - \frac{23302711005\,\ep^{19}}{2}\,\zeta_{22} + \frac{1194695479813\,\ep^{21}}{2}\,\zeta_{24} - \frac{72628776062025\,\ep^{23}}{2}\,\zeta_{26} + 
\nonum \\
&+ \frac{5165901157067001\,\ep^{25}}{2}\,\zeta_{28} - \frac{425013158488292213\,\ep^{27}}{2}\,\zeta_{30} + \frac{40048437741888549165\,\ep^{29}}{2}\,\zeta_{32} + \Ord(\ep^{31})\, ,
\\
\hat{\zeta}_{5} &= \boxed{\zeta_5 + \frac{5\,\ep}{2}\,\zeta_6 - \frac{35\,\ep^3}{4}\,\zeta_8 + 63\,\ep^5\,\zeta_{10}}
~	\scalebox{1}{\dbox{$- \frac{2805\,\ep^7}{4}\,\zeta_{12}$}} +
\frac{22165\,\ep^9}{2}\,\zeta_{14} - \frac{943215\,\ep^{11}}{4}\,\zeta_{16} + 6498590\,\ep^{13}\,\zeta_{18} - 
\nonum \\
&- \frac{900752361\,\ep^{15}}{4}\,\zeta_{20} + \frac{19165711635\,\ep^{17}}{2}\,\zeta_{22} - \frac{1965195294755\,\ep^{19}}{4}\,\zeta_{24} + 29867386995325\,\ep^{21}\,\zeta_{26} -
\nonum \\
&- \frac{8497566799256925\,\ep^{23}}{4}\,\zeta_{28} + \frac{349559311628200401\,\ep^{25}}{2}\,\zeta_{30} - \frac{65877039565685293015\,\ep^{27}}{4}\,\zeta_{32} + \Ord(\ep^{29})\, ,  
\\
\hat{\zeta}_{7} &= \boxed{\zeta_7 + \frac{7\,\ep}{2}\,\zeta_8 - 21\,\ep^3\,\zeta_{10}}
	~\scalebox{1}{\dbox{$+ 231\,\ep^5\,\zeta_{12}$}} - \frac{7293\,\ep^7}{2}\,\zeta_{14} +
\frac{155155\,\ep^9}{2}\,\zeta_{16} - 2137954\,\ep^{11}\,\zeta_{18} + 74083926\,\ep^{13}\,\zeta_{20} -
\nonum \\
&-\frac{6305266527\,\ep^{15}}{2}\,\zeta_{22} + \frac{323261669577\,\ep^{17}}{2}\,\zeta_{24} - 9825976473775\,\ep^{19}\,\zeta_{26} + 698896855690605\,\ep^{21}\,\zeta_{28} -
\nonum \\
&- \frac{115000404016610385\,\ep^{23}}{2}\,\zeta_{30} + \frac{10836338660474212431\,\ep^{25}}{2}\,\zeta_{32} - 579717948178030578532\,\ep^{27}\,\zeta_{34} + \Ord(\ep^{29})\, ,
\\
\hat{\zeta}_{9} &= \boxed{\zeta_9 + \frac{9\,\ep}{2}\,\zeta_{10}}
	~\scalebox{1}{\dbox{$- \frac{165\,\ep^3}{4}\,\zeta_{12}$}} + \frac{1287\,\ep^5}{2}\,\zeta_{14} - \frac{109395\,\ep^7}{8}\,\zeta_{16} +
376805\,\ep^9\,\zeta_{18} - \frac{26113581\,\ep^{11}}{2}\,\zeta_{20} + 555629445\,\ep^{13}\,\zeta_{22} -
\nonum \\
&-\frac{227890347333\,\ep^{15}}{8}\,\zeta_{24} + \frac{3463517888325\,\ep^{17}}{2}\,\zeta_{26} - \frac{492702534613575\,\ep^{19}}{4}\,\zeta_{28} + \frac{20268008815027545\,\ep^{21}}{2}\,\zeta_{30} -
\nonum \\
&- \frac{7639312552531975575\,\ep^{23}}{8}\,\zeta_{32} + 102171193084471145778\,\ep^{25}\,\zeta_{34} - 12319006398783149793805\,\ep^{27}\,\zeta_{36} + \Ord(\ep^{29})\, ,
\\
	\hat{\zeta}_{11} &= \boxed{\zeta_{11}}~\scalebox{1}{\dbox{$+ \frac{11\,\ep}{2}\,\zeta_{12}$}} - \frac{143\,\ep^3}{2}\,\zeta_{14} + \frac{3003\,\ep^5}{2}\,\zeta_{16} - 41327\,\ep^7\,\zeta_{18} 
+ 1431859\,\ep^9\,\zeta_{20} - 60931689\,\ep^{11}\,\zeta_{22} + 3123872213\,\ep^{13}\,\zeta_{24} -
\nonum \\
&-\frac{379817245555\,\ep^{15}}{2}\,\zeta_{26} + \frac{27015439528935\,\ep^{17}}{2}\,\zeta_{28} - \frac{2222635878367905\,\ep^{19}}{2}\,\zeta_{30} + \frac{209436091088617965\,\ep^{21}}{2}\,\zeta_{32} -
\nonum \\
&- 11204325077046897510\,\ep^{23}\,\zeta_{34} + 1350930219672451816398\,\ep^{25}\,\zeta_{36} + \Ord(\ep^{27})\, ,
\\
	\hat{\zeta}_{13} &= \scalebox{1}{\dbox{$\zeta_{13}$}} + \frac{13\,\ep}{2}\,\zeta_{14} - \frac{455\,\ep^3}{4}\,\zeta_{16} + 3094\,\ep^5\,\zeta_{18} - \frac{214149\,\ep^7}{2}\,\zeta_{20}
+ 4555915\,\ep^9\,\zeta_{22} - \frac{467142949\,\ep^{11}}{2}\,\zeta_{24} + 14199419150\,\ep^{13}\,\zeta_{26} -
\nonum \\
&-\frac{4039874339085\,\ep^{15}}{4}\,\zeta_{28} + \frac{166185885587085\,\ep^{17}}{2}\,\zeta_{30} - \frac{31318960104275025\,\ep^{19}}{4}\,\zeta_{32} + 837744364354471860\,\ep^{21}\,\zeta_{34} -
\nonum \\
&- 101008688194589454825\,\ep^{23}\,\zeta_{36} + \Ord(\ep^{25})\, ,
\\
\hat{\zeta}_{15} &= {\zeta_{15}} + \frac{15\,\ep}{2}\,\zeta_{16} - 170\,\ep^3\,\zeta_{18} + 5814\,\ep^5\,\zeta_{20} - 247095\,\ep^7\,\zeta_{22}
+ 12666445\,\ep^9\,\zeta_{24} - 770015850\,\ep^{11}\,\zeta_{26} + 
\nonum \\
&+ 54769188150\,\ep^{13}\,\zeta_{28} - \frac{9012027371805\,\ep^{15}}{2}\,\zeta_{30} + \frac{849191613164775\,\ep^{17}}{2}\,\zeta_{32} - 45429700371036300\,\ep^{19}\,\zeta_{34} + 
\nonum \\
& + 5477559305394623700\,\ep^{21}\,\zeta_{36} + \Ord(\ep^{23})\, ,
\\
\hat{\zeta}_{17} &= \zeta_{17} + \frac{17\,\ep}{2}\,\zeta_{18} - \frac{969\,\ep^3}{4}\,\zeta_{20} + \frac{20349\,\ep^5}{2}\,\zeta_{22} - \frac{4167669\,\ep^7}{8}\,\zeta_{24}
+ \frac{63332225\,\ep^9}{2}\,\zeta_{26} - \frac{9009185445\,\ep^{11}}{4}\,\zeta_{28} +
\nonum \\
&+ \frac{370604839815\,\ep^{13}}{2}\,\zeta_{30} - \frac{279372848525955\,\ep^{15}}{16}\,\zeta_{32} + 1868221548962505\,\ep^{17}\,\zeta_{34} - \frac{450511195346109975\,\ep^{19}}{2}\,\zeta_{34} +
\nonum \\
& + \Ord(\ep^{21})\, ,
\\
\hat{\zeta}_{19} &= \zeta_{19} + \frac{19\,\ep}{2}\,\zeta_{20} - \frac{665\,\ep^3}{2}\,\zeta_{22} + \frac{33649\,\ep^5}{2}\,\zeta_{24} - \frac{2042975\,\ep^7}{2}\,\zeta_{26}
+ \frac{145291575\,\ep^9}{2}\,\zeta_{28} - \frac{11953363695\,\ep^{11}}{4}\,\zeta_{30} +
\nonum \\
&+ \frac{1126348042575\,\ep^{13}}{2}\,\zeta_{32} - 60256888897755\,\ep^{15}\,\zeta_{34} + 7265306023743075\,\ep^{17}\,\zeta_{36} + \Ord(\ep^{19})\, .
\end{flalign}
\end{subequations}

\end{widetext}

\end{document}